\documentstyle{camera}

\title{BeppoSAX/PDS OBSERVATION OF VELA X--1}
\author{M. Orlandini, D. Dal~Fiume, L. Nicastro, \\ 
        E. Palazzi \And F. Frontera\ddag}
\organization{Istituto TeSRE/CNR Bologna \\
              \ddag\ also Dipartimento di Fisica, Universit\`a di Ferrara}

\def\reference{\noindent\hangindent=20pt\hangafter=1}
\def\V{Vela X--1}   \def\B{BeppoSAX}

\begin{document}

\maketitle

\begin{abstract}
\noindent Vela X--1 has been observed by the Italian/Dutch X--ray satellite
BeppoSAX during its Science Verification Phase. During the $\sim 50$ ksec
observation (orbital phases 0.27--0.34) two different intensity state are
present, situation very common in this source. We  report results from the
pulse phase averaged and pulse phase resolved spectroscopy from the high-energy
instrument PDS aboard \B.
\end{abstract}

\section{Introduction}

The X--ray binary system \V\ is composed by a neutron star and the B0.5
supergiant HD77851 (Hutchings 1974). The orbit is almost circular, with an
orbital period of 8.96 days. The neutron star pulses at about 283 seconds
(McClintock et~al.\ 1976), and is occulted by the companion for 1.7 days
(Avni 1976). The contemporary presence of pulsations and X--ray eclipses
has allowed the determination of the geometry of the system: the neutron star
is heavily embedded in the strong stellar wind and both accretion and
photoionization wakes are formed during the orbital motion of the neutron star
(Kaper et~al.\ 1997).

\V\ shows large temporal variability on time scales from minutes to hours. They
have been ascribed to fluctuations in the accretion rate and to inhomogeneities
in the stellar wind (Haberl \& White 1990). The pulse period itself has changed
during the more than 20 years of observations: the variations follows a wavy
behaviour, due to fluctuations in the transfer of angular momentum from the
wind to the neutron star. The observed flip-flop behaviour of the \V\ spin
history has been explained as due to the formation of a temporary accretion
disk, that acts as reservoir of angular momentum (B{\"o}rner et~al.\ 1987). The
problem with this interpretation is the very high magnetic field strength
necessary for the disk formation.

The X--ray spectrum of \V\ has been described by the usual (for X--ray pulsars)
power law modified by an exponential cutoff at high energies (White et~al.\
1983). Due to the absorption of the wind, the pulse-averaged spectrum is very
variable along the orbit (Choi et~al.\ 1996): this has been used as tracer of
the circumstellar matter. The spectrum also shows an iron emission line (Becker
et~al.\ 1978) and line features at $\sim 27$ and $\sim 54$ keV that have been
interpreted as cyclotron absorption lines (Makishima \& Mihara 1992; Kretschmar
et~al.\ 1996).

\section{Observation}

The \B\ satellite is a joint program of the Italian Space Agency (ASI) and the
Netherlands Agency for Aerospace Programs (NIVR).  The payload includes Narrow
Field Instruments (NFI) and Wide  Field Cameras (WFC). The NFIs are four
Concentrators Spectrometers (C/S) with 3 units (MECS) operating in the 1--10
keV energy band (Boella et~al.\ 1997) and 1 unit (LECS) operating in 0.1--10
keV (Parmar et~al.\ 1997), a High Pressure Gas Scintillation Proportional
Counter (HPGSPC) operating in the 3--120 keV energy band (Manzo et~al.\ 1997)
and a Phoswich Detection System (PDS) with four detection units operating in
the 15--300 keV energy band (Frontera et~al.\ 1997). Orthogonally with respect
to the NFIs there are two WFCs (field of view of $20^\circ \times 20^\circ$
FWHM) which operate in the 2--30 keV energy band with imaging capabilities
(angular resolution of 5 arcmin) (Jager et~al.\ 1997).

During the Science Verification Phase (SVP) a series of known X--ray sources
have been observed in order to check the capabilities and performances of the
instruments aboard \B. \V\ is one of these sources and it has been observed by
three of the four NFIs (LECS was not operative during this pointing).

\begin{figure}
\vspace{6cm}
\includegraphics{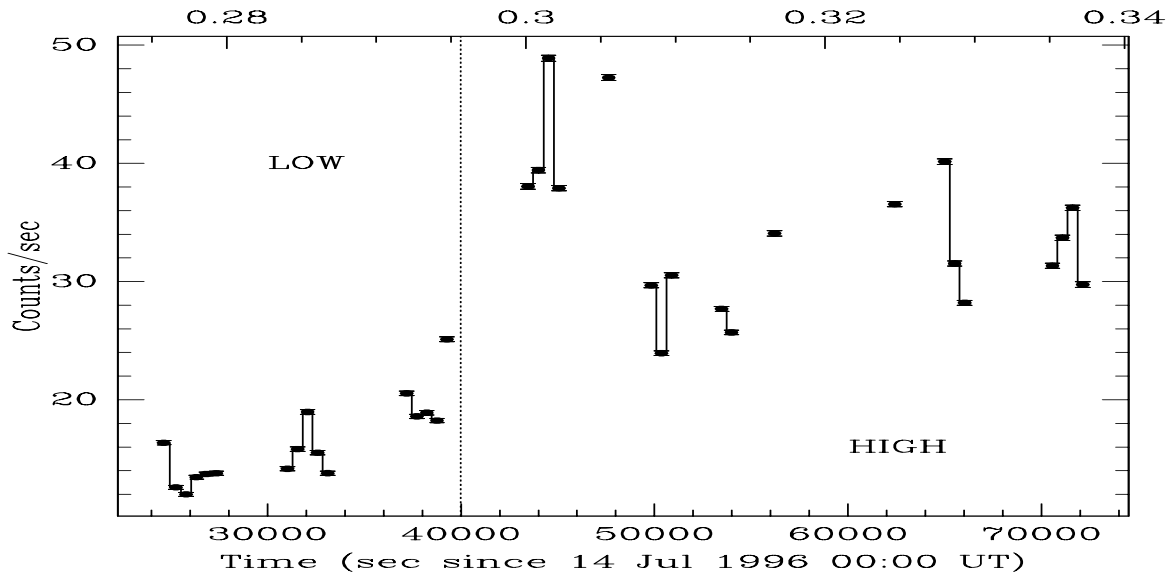}
\caption[]{\V\ 500 sec background subtracted light curve for 15--300 keV
SAX/PDS data. Gaps are due to South Atlantic Anomaly passages, where the
instruments are switched off, and occultations by the Earth.}
\label{light_curve}
\end{figure}

In Fig.~\ref{light_curve} we show the 15--300 keV background subtracted light
curve of \V\ has observed by PDS. The operative mode during the
observation was DIR001, corresponding to the maximum time and energy resolution
(15 $\mu$sec and 1024 energy channels, respectively).

The upper scale represents the orbital phase referred to the ephemeris given by
Deeter et~al.\ (1987). The first part of the observation (marked LOW in the
figure) corresponds to one of the common intensity dips showed by \V. Its origin
is due to the clumpsy circumstellar material that intercepts the emission
coming from the pulsar (Sato et~al.\ 1986).

The background evaluation has been performed using the rocking collimator
capability of PDS, that allows the contemporary monitoring of the source and
the background. The dwell time for this observation was 50 sec, and a Fourier
analysis of the background time series did not show the presence of spurious
frequencies due to the rocking.

The data analysis has been performed with the XAS package, version 2.0.1,
developed by Lucio Chiappetti.

\begin{figure}
\vspace{6cm}
\includegraphics{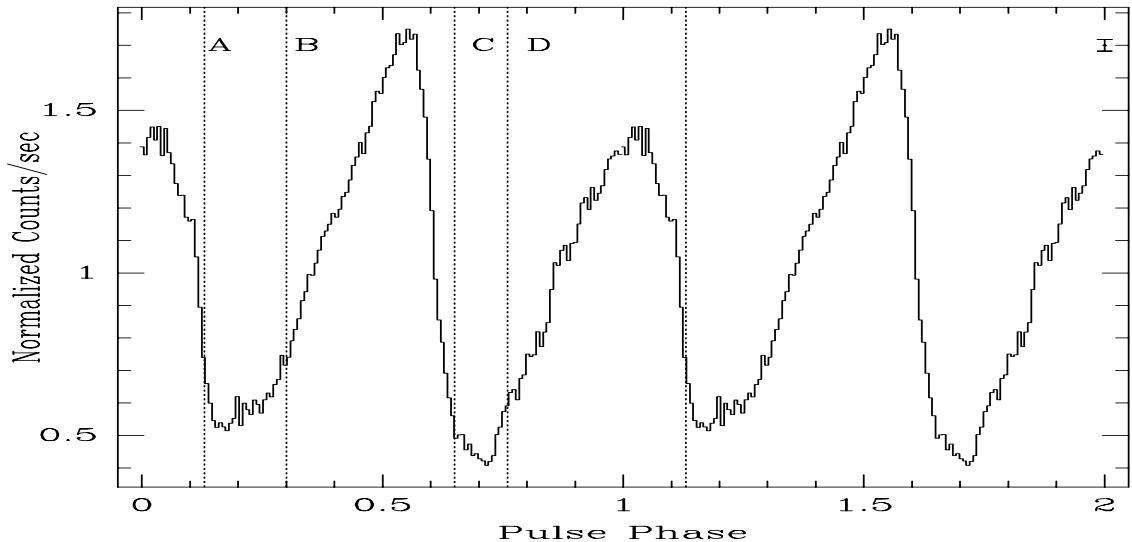}
\caption[]{15--300 keV \V\ normalized pulse profile, together with pulse phase
intervals chosen for the pulse phase spectroscopy.}
\label{pulse}
\end{figure}

\section{Data Analysis}

In Fig.~\ref{pulse} we present the 15--300 keV pulse profile of \V, showing the
characteristic double peak structure typical in this energy range (Orlandini
1993). We performed pulse phase spectroscopy on the \V\ PDS data by choosing
four pulse phase intervals, as described in the same figure. The 20--200 keV
data has been fit with a power law plus exponential cutoff, with the addition
of two cyclotron absorption lines (Mihara et~al.\ 1990) for the pulse peak
spectra. The results are summarized in Table~\ref{fit_results}.

\begin{table}
\begin{center}
{\footnotesize\begin{tabular}{c|c|c|c|c|c} \hline
\multicolumn{1}{c|}{Pulse Phase} & Avg & A & B & C & D \\ \cline{1-1}
\multicolumn{1}{c|}{Fit Param} & & & & & \\ \hline\hline
$\alpha$   & $2.19\pm 0.02$ & $2.44\pm 0.05$ & $2.23\pm 0.02$ & $2.42\pm 0.06$ & $2.25\pm 0.02$ \\
$I_{70}^\ast$& $3.2\pm 0.2$ & $2.1\pm 0.4$   & $4.5\pm 0.4$   & $2.5\pm 0.6$   & $4.4\pm 0.4$ \\
$E_c$ (keV)& $33.3\pm 0.1$  & $32.8\pm 0.5$& $35.6\pm 0.2$  & $33.8\pm 0.6$  & $32.7\pm 0.2$ \\
$E_f$ (keV)& $11.9\pm 0.1$  & $12.8\pm 0.5$& $12.4\pm 0.2$  & $13.4\pm 0.7$  & $13.0\pm 0.2$ \\
$\chi^2_{\rm dof}$ & 4.81 & 1.21 & 4.23           & 1.07           & 4.71 \\
\hline\hline
$\alpha$   & $1.85\pm 0.09$ &             & $1.45\pm 0.15$   & & $1.73\pm 0.14$ \\
$I_{70}^\ast$& $7\pm 2$     &             & $7\pm 3$         & & $15\pm 8$ \\
$E_c$ (keV)& $30.7\pm 0.5$  &             & $30.6\pm 0.7$    & & $29.4\pm 0.4$ \\
$E_f$ (keV)& $14.3\pm 0.8$  &             & $12\pm 1$        & & $19\pm 2$ \\
$E_{\rm cyc}$ (keV) & $28.1\pm 0.2$ &     & $28.3\pm 0.3$    & & $28.9\pm 0.4$ \\
$\tau_1\ ;\ W_1$ (keV) & $0.19\pm 0.06\ ;\ 0$ & & $0.28\pm 0.07\ ;\ 5\pm 2$ & & 
                            $0.5\pm 2\ ;\ 0.2\pm 1$ \\
$\tau_2\ ;\ W_2$ (keV) & $0.8\pm 0.1\ ;\ 9\pm 2$ & & $1.2\pm 0.2\ ;\ 5\pm 2$ & & 
                           $1.2\pm 0.2\ ;\ 17\pm 3$ \\
$\chi^2_{\rm dof}$ & 2.08 &     & 0.93             & & 1.46 \\ \hline
\multicolumn{6}{l}{$^\ast$ Flux at 70 keV in $10^{-5}$ ph~cm$^{-2}$~sec$^{-1}$} \\
\end{tabular}
}
\end{center}
\caption[]{Results of the fits to the 20--200 keV \V\ phase resolved spectra
with a power law plus exponential cutoff (upper panel), and the same continuum
plus absorption cyclotron lines (lower panel). $\tau$ and $W$ corresponds to
the depth and half width of the line, respectively. The energy of the first
harmonic is fixed as the double of the fundamental.}
\label{fit_results}
\end{table}

The 20-200 keV flux is $5.3\times 10^{-9}$ erg~cm$^{-2}$~s$^{-1}$,
corresponding to an average X--ray luminosity of $2.3\times 10^{36}$ erg/sec.

As is evident from the Table~\ref{fit_results}, the inclusion of cyclotron
absorption improves the fits for spectra taken in the pulses, while it is not
necessary in the valley spectra. Actually, this is only due to poor statistics,
because a ratio between spectra taken in the pulses and in the valleys is
nearly constant up to 40 keV. The cyclotron resonance energies in the two
pulses are consistent with each other, but in phase D (the secondary peak) only
the second harmonic is required. Note also a hardening of the spectrum in the
main peak.

We also performed a ratio between the spectra and the featureless spectrum from
the X--ray pulsar Crab, in order to reduce systematic errors. The fit still
requires the inclusion of a cyclotron resonance, with energy of the fundamental
at $E_{\rm cyc} = 28.5\pm 0.2$ keV for the average spectrum.


\small\section*{References}

\reference{Avni, Y. 1976, ApJ, 209, 574}

\reference{Becker, R.H., et~al.\ 1978, ApJ, 221, 912}

\reference{Boella, G., et~al.\ 1997, A\&A, 320, in press}

\reference{B{\"o}rner, G., et~al.\ 1987, A\&A, 182, 63}

\reference{Choi, C.S., et~al.\ 1996, ApJ, 471, 447}

\reference{Deeter, J.E., et~al.\ 1987,  AJ, 93, 877}

\reference{Frontera, F., et~al.\ 1997, A\&A, 320, in press}

\reference{Haberl, F., \& White, N.E. 1990, ApJ, 361, 225}

\reference{Hutchings, J.B. 1974, ApJ, 192, 685}

\reference{Jager, R., et~al.\ 1997, A\&A, 320, in press}

\reference{Kaper, L., et~al.\ 1997, ApJ, 475, L37}

\reference{Kretschmar, P., et~al.\ 1996, A\&AS, 120, 175}

\reference{Makishima, K., \& Mihara, T. 1992,
 in Frontiers of X--ray Astronomy, eds. Tanaka, Y., \& Koyama, K.
  {Universal Academy Press, Tokyo}, 23}

\reference{Manzo, G., et~al.\ 1997, A\&A, 320, in press}

\reference{McClintock, J.E., et~al.\ 1976, ApJ, 206, L99}

\reference{Mihara, T., et~al.\ 1990, Nat, 346, 250}

\reference{Orlandini, M. 1993, MNRAS, 264, 181}

\reference{Parmar, A.N., et~al. 1997, A\&A, 320, in press}

\reference{Sato, N., et~al.\ 1986, PASJ, 38, 731}

\reference{White, N.E., Swank, J.H., \& Holt, S.S. 1983,  ApJ, 270, 711}

\end{document}